\documentclass{optica-article}

\journal{opticajournal} 

\articletype{Research Article}

\usepackage{lineno}

\begin{document}

\title{Simulation of Speckle Interferometric Results for Enhanced Measurement and Automated Defect Detection
}

\author{Jessica Plaßmann\authormark{1,2,*}, Michael Schuth\authormark{2} and Georg von Freymann\authormark{1,3}}

\address{\authormark{1}Department of Physics and research center OPTIMAS, RPTU University Kaiserslautern-Landau, 67663 Kaiserslautern, Germany\\
\authormark{2}Department of Engineering and technology center OGKB, Trier University of Applied Sciences, Schneidershof 1, 54293 Trier, Germany\\
\authormark{3}Fraunhofer Institute for Industrial Mathematics ITWM, 67663 Kaiserslautern, Germany}

\email{\authormark{*}plassmaj@hochschule-trier.de} 


\begin{abstract*} 

Techniques like speckle holography and shearography are rarely applied due to the complexity of instrument setup and lack of automated result analysis, despite their potential. By simulating speckle interferometric outcomes, we seek to address these challenges, enabling more efficient measurement processes and paving the way for automated defect recognition. This research focuses on developing a simulation code for speckle interferometric results derived from finite element analyses. The aim is to improve the parameter settings of speckle interferometry measurements and create specific datasets, which will be used to develop machine learning-based methods for automation in series production.
\end{abstract*}

\section{Introduction} 

Shearography, highlighted in 1980 by Nakadate et al. \cite{Nakadate:80}, has become a widely recognized non-contact and non-destructive testing (NDT) method. It is capable of measuring strain changes in the micrometer range and is particularly suitable for detecting defects in applications where conventional methods reach their physical or practical limits \cite{Petry:21}. Over the past decades, significant advancements in shearography technology have been made. For instance, dual-shearing systems \cite{Jiang:20} and methods employing spatial light modulators (SLMs) \cite{Zhang:20,Wang:23} have greatly improved the technique's versatility and efficiency. A recently introduced high-performance rotational shearography system based on a Dove prism further enhances defect detection by enabling simultaneous measurements with a single temporal phase shift, minimizing missed detections even in large or small fields of view \cite{Yao:25}. 
These innovations enable simultaneous measurements of deformation derivatives in multiple directions and facilitate faster and more flexible imaging setups, particularly for complex material testing. 

Despite these advancements, certain challenges remain that hinder the widespread adoption of shearography, particularly in industrial serial production. One major limitation is the optimization of testing parameters, which include loading conditions and device-specific settings. Currently, the selection of these parameters is highly dependent on the experience of inspectors, which limits the reproducibility and economic efficiency of the method. Additionally, while shearography is highly effective in detecting localized defects, ensuring that the chosen parameters reliably reveal all critical defects in a given component remains a significant challenge. \cite{Plassmann:21}

Simulation presents a promising approach to address these issues. By simulating surface deformations under various loading conditions and translating them into shearographic images, it is possible to systematically optimize the testing parameters before performing physical measurements. Previous studies have demonstrated the potential of simulation-based approaches to enhance the accuracy and efficiency of shearographic testing, especially in cases involving complex materials or testing environments. \cite{Tao:22a, Tao:22b}

Studies like Li (2022) \cite{Li:22} have highlighted the potential of combining real and synthetic data to improve defect detection in shearography. However, their findings also reveal that overly diverse synthetic datasets can lead to overfitting, particularly when applied to larger datasets. This underscores the importance of tailoring datasets to specific components and testing conditions. Although dataset generation is not the central objective of this work, the proposed simulation framework provides a foundation for creating component-specific datasets, enhancing the reliability and applicability of hybrid training approaches in future research. Such datasets could alleviate the challenges associated with the time-consuming and complex process of manually creating large, high-quality datasets, while also improving the adaptability and generalizability of machine learning models. By augmenting and refining datasets synthetically, the applicability and reliability of defect detection methods can be greatly enhanced, particularly in industrial serial production.

The central goal of this work is therefore the development of a validated and openly accessible code base, focused initially on simulating and validating out-of-plane deformation. In subsequent research, the extension to include in-plane deformation is planned. The source code will be made available in a dedicated repository at \cite{Plassmann:25}, enabling transparent collaboration and further development. 

\section{Method}
\subsection{Design of Test Component} \label{sssec:defect}
A thorough literature review on shearographic testing revealed a diverse range of materials under investigation. Isotropic materials \cite{Yang:17, Wang:19}, including aluminum (24$\%$), polymers, and glass, were prominently featured. Anisotropic materials were also well-represented, with glass fiber-reinforced composites (GFRP, 19$\%$) \cite{Yang:17, Maranon:07} and carbon fiber-reinforced composites (CFRP, 14$\%$) \cite{Tao:22a, Tao:22b} being notable examples.

This study strategically focuses on isotropic materials for validation under well-controlled and predictable conditions. By selecting a material with uniform mechanical properties, it becomes possible to isolate testing parameter effects from material-specific behavior. This approach enables a rigorous evaluation of the framework's capabilities before extending to more complex anisotropic materials. Polyamide 6 (PA6) was chosen as the test material, owing to its isotropic properties and significant industrial relevance.

\begin{figure}[htbp]
\centering\includegraphics[height=3.5cm]{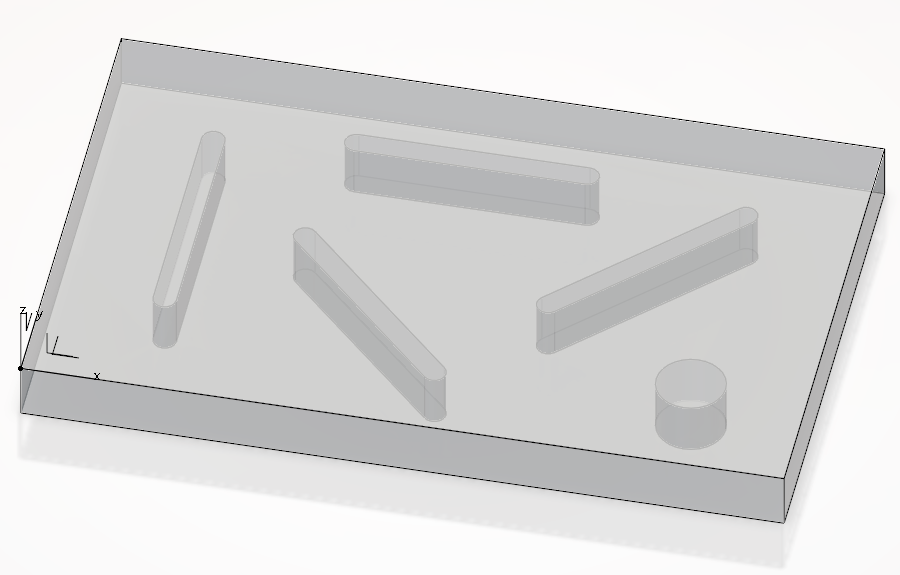}
\centering\includegraphics[height=3.5cm, trim=4cm 4cm 4cm 4cm, clip]{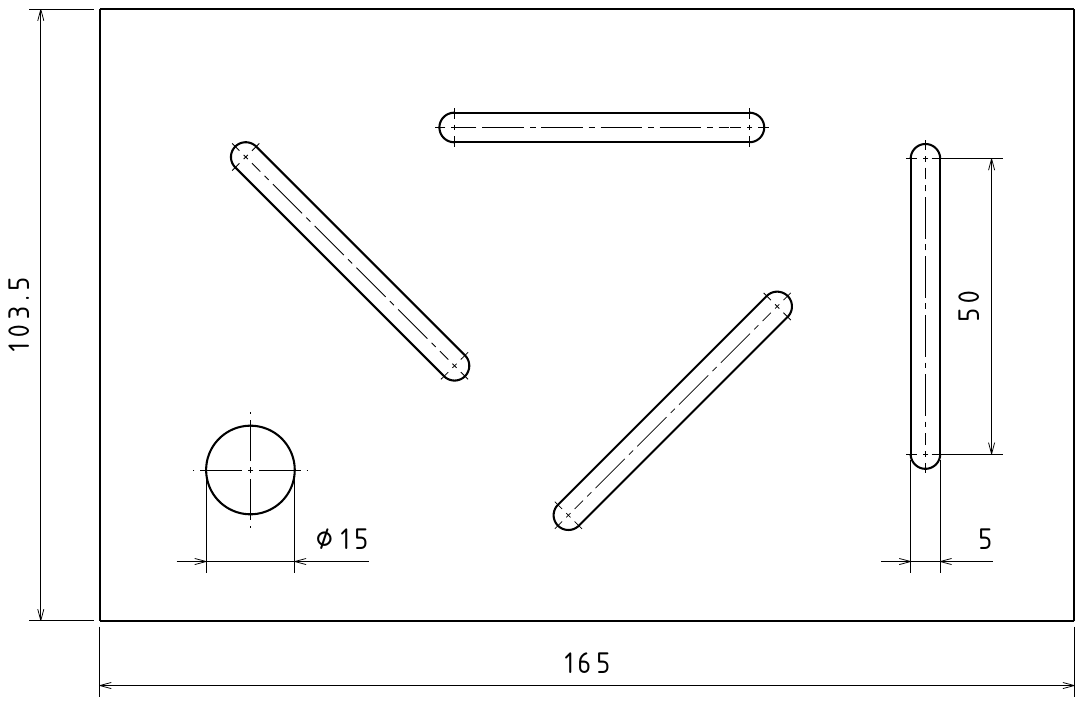}
\caption{Designed test component with applied defects (left) and corresponding sketch showing main dimensions (right)}
\label{fig:geometry}
\end{figure}

The literature review also identified the predominant defect types in shearographic research, with flat-bottom holes (36$\%$) \cite{Tao:22a, Tao:22b, Li:22}, cracks (22$\%$) \cite{Wen:2012}, and notches (21$\%$) \cite{Buchta:17} being the most common. The test component was prepared with flat-bottom holes and cracks of varying orientations to simulate realistic defect scenarios. Figure \ref{fig:geometry} illustrates the bottom view of the designed test component, showcasing the defects applied. This setup serves to validate the simulation code, which is intended to be transferable to other geometries in future work.

\subsection{Loading Methods}
Shearography employs various loading methods to induce surface deformations in test objects, which are necessary for identifying internal or subsurface defects. The selection of an appropriate loading method depends on the material properties, defect types, and the configuration of the object under investigation. Different phase-shifting techniques used in shearographic systems further influence the applicability of these loading methods, as they enable both static and dynamic measurements depending on the system setup.

Thermal loading introduces heat into the object, leading to expansion. Variations in thermal conductivity or expansion coefficients between defective and intact regions can result in local deformation differences. Under- or overpressure methods change the internal or external pressure of the object, generating stresses particularly suitable for evaluating hollow or bonded structures. Mechanical loading applies localized force to the object’s surface or structure, while vibrational excitation uses oscillatory input — commonly from piezoelectric actuators — to identify defects that respond to dynamic stimulation. Each method produces specific deformation patterns that can enhance defect visibility under shearographic inspection.

An analysis of recent studies shows a diverse spectrum of loading methods employed in the field. Thermal loading was used in 35$\%$ of the examined cases \cite{Yang:17, Tao:22a, Tao:22b, Buchta:17}. Under- or overpressure methods accounted for 29$\%$ \cite{Yang:17, Maranon:07}, while mechanical force application and vibrational excitation were each reported in 18$\%$ of the cases \cite{Yang:17, Wang:19, Liu:18}. This distribution indicates a preference for thermal and pressure-based techniques in many experimental and practical contexts.

Based on this literature review, thermal loading using a halogen lamp was selected for the present study. This approach enables contactless excitation, which reduces the influence of external mechanical constraints on the measurement. The heat generated by the halogen source can be distributed even across the surface, and the thermal input can be adjusted with respect to intensity and duration. The method's versatility makes it suitable for a wide range of materials and structures, while being more cost-efficient compared to complex mechanical or pneumatic systems. The selection of thermal loading is further supported by advances in spatial phase-shifting techniques, which allow interference phase determination in a single image acquisition per load condition. This improves measurement stability and enables dynamic detection of impact damages, cracks, and delaminations \cite{Petry:20}.

\subsection{Simulation Framework}

\subsubsection{Finite Element Analysis Model}

A coupled thermomechanical transient analysis was conducted in ANSYS Workbench 2024 R2 to evaluate the temporal evolution of deformation behavior under thermal excitation. The simulation framework considers both the heating and cooling phases to capture the full thermomechanical response of the specimen. Thermal loading was defined as a spatially uniform surface heat flux with a total duration of 1.6~s, starting at $t~=~0~\text{s}$. The excitation profile included a ramp-up phase of 0.1~s, a constant heating period of 1.4~s, and a ramp-down phase of 0.1~s. The heat flux profile was validated by matching the simulated surface temperature evolution to experimental data acquired through infrared thermography using the FLIR A65 thermographic camera.

The finite element model included the definition of temperature-dependent material properties, geometric representation of the specimen and structured meshing. Mechanical boundary conditions were implemented based on the experimental fixture. The specimen was placed onto a support with known thermal conductivity, and the surface in contact with the support was constrained in the direction normal to this surface. A mesh independence study was conducted to ensure numerical stability and convergence of deformation results.

\subsubsection{Generation of Synthetic Shearograms}
In shearography, the simulation of measurement results requires specific displacement gradients derived from the finite element analysis (FEA). For the out-of-plane technique, the gradients \(\frac{\partial w}{\partial x}\) and \(\frac{\partial w}{\partial y}\) are key quantities, where $w$ represents the displacement in the $z$-direction (normal to the surface). 
For the in-plane technique, the relevant quantities are the gradients \(\frac{\partial u}{\partial x}\), \(\frac{\partial u}{\partial y}\), \(\frac{\partial v}{\partial x}\), and \(\frac{\partial v}{\partial y}\), where $u$ and $v$ represent displacements along the $x$- and $y$-directions, respectively. 
These gradients are directly linked to components of the strain tensor $\varepsilon$, which for small deformations can be expressed as shown in Eq. (\ref{eq:straintensor}):

\begin{equation}\label{eq:straintensor}
\varepsilon =
\begin{bmatrix}
\varepsilon_{xx} & \frac{1}{2} \gamma_{xy} & \frac{1}{2} \gamma_{xz} \\
\frac{1}{2} \gamma_{yx} & \varepsilon_{yy} & \frac{1}{2} \gamma_{yz} \\
\frac{1}{2} \gamma_{zx} & \frac{1}{2} \gamma_{zy} & \varepsilon_{zz}
\end{bmatrix}
=
\begin{bmatrix}
\frac{\partial u}{\partial x} & \frac{1}{2} \left( \frac{\partial u}{\partial y} + \frac{\partial v}{\partial x} \right) & \frac{1}{2} \left( \frac{\partial u}{\partial z} + \frac{\partial w}{\partial x} \right) \\
\frac{1}{2} \left( \frac{\partial u}{\partial y} + \frac{\partial v}{\partial x} \right) & \frac{\partial v}{\partial y} & \frac{1}{2} \left( \frac{\partial v}{\partial z} + \frac{\partial w}{\partial y} \right) \\
\frac{1}{2} \left( \frac{\partial u}{\partial z} + \frac{\partial w}{\partial x} \right) & \frac{1}{2} \left( \frac{\partial v}{\partial z} + \frac{\partial w}{\partial y} \right) & \frac{\partial w}{\partial z}
\end{bmatrix}
\end{equation}

To simulate shearographic measurement results, the FEA displacement data are transformed into phase maps using the general formula for the shearographic fringe order \(N\) as described in \cite[p.~39]{Petry:21}:

\begin{equation}\label{eq:fringe}
N = \frac{\Delta_{\text{ESPSI}_{\Theta_{xz}}}} {2\pi} = \frac{\delta_x}{\lambda} 
\left[ \frac{\partial u}{\partial x} \sin(\Theta_{xz}) + \frac{\partial w}{\partial x} \left( 1 + \cos(\Theta_{xz}) \right) \right]
\end{equation}

In Eq. (\ref{eq:fringe}), \(N\) represents the shearographic fringe order, and \(\Delta_{\text{ESPSI}_{\Theta_{xz}}}\) is the phase change in digital shearography. The parameter \(\lambda\) denotes the laser wavelength, while \(\Theta_{xz}\) is the illumination angle in the \(xz\)-plane. The displacement components \(u\) and \(w\) correspond to the in-plane displacement in the \(x\)-direction and the out-of-plane displacement in the \(z\)-direction, respectively. The shear amount in the \(x\)-direction is given by \(\delta_x\), with \(\frac{\partial u}{\partial x}\) representing the in-plane strain in the \(x\)-direction and \(\frac{\partial w}{\partial x}\) representing the out-of-plane slope in the \(x\)-direction.

For the out-of-plane technique, \(\Theta_{xz}\) approaches $0^\circ$, which leads to the simplification of the formula as follows:
\begin{equation}\label{eq:Noop}
N_{\text{oop}} = \frac{\Delta_{\text{ESPSI}_{\Theta_{xz}}}}{2\pi} = \frac{2 \delta_x}{\lambda} \frac{\partial w}{\partial x}
\end{equation}


Using Eq. (\ref{eq:Noop}) 
, the FEA-derived displacement gradients are converted into phase maps using MATLAB, which represent the interferometric phase changes across the surface of the test object. The phase maps are wrapped within the interval $\left[ 0, 2\pi \right]$, introducing phase discontinuities (jumps) that are characteristic of shearographic measurements. These phase maps are computed based on the optical parameters of the system, including the laser wavelength, shearing distance, and illumination angle, as well as the specific characteristics of the camera utilized during the measurement, ensuring an accurate digital representation of the experimental setup.

This simulation establishes a reference scenario under idealized conditions, in which device parameters are modeled accurately and neither noise nor speckle patterns are included. The resulting data serve to evaluate the fundamental sensitivity of the shearographic method. In subsequent studies, this configuration will be extended to systematically assess the influence of measurement noise, image processing techniques, and parameter variations on defect detection performance. 

\subsection{Validation Process - Experimental Shearographic Testing}

\begin{figure}[htbp]
    \centering
    \includegraphics[width=0.8\linewidth]{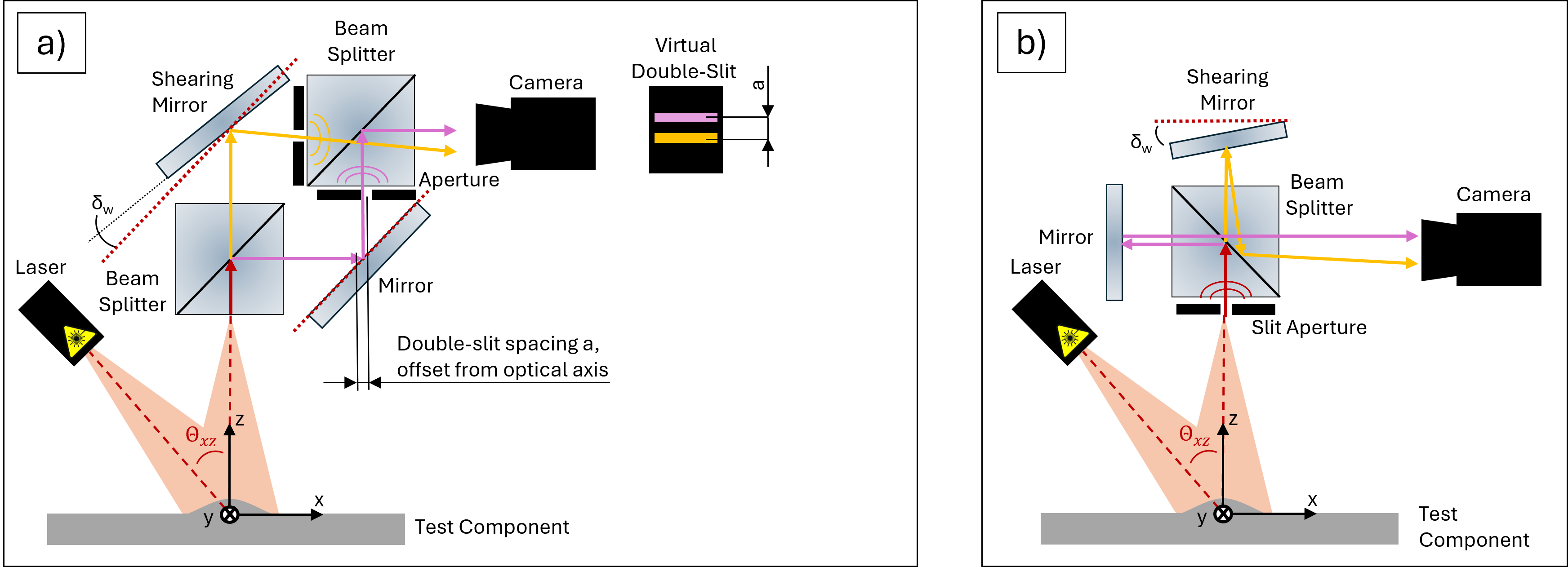}
    \caption{Schematic representation of shearography measurement setups using spatial phase shifting: (a) Mach–Zehnder interferometer and (b) Michelson interferometer. Own illustration based on \cite[p. 86, 97]{Petry:21}}
    \label{fig:setups}
\end{figure}

The experimental validation of the simulation framework was conducted using two custom-built shearography systems, employing Michelson and Mach-Zehnder interferometer configurations (see Fig. \ref{fig:setups}). Both setups were designed to closely match the parameters used in the simulation, ensuring a direct comparison between experimental and simulated results. A key feature of both interferometer setups was the adjustable shear distance, varied between 1~mm and 5~mm. This allowed investigation of the shear distance's influence on defect detectability and enabled validation of the simulation under different shearing conditions. For phase-shifting, a spatial phase-shifting method with a virtual double slit was used. This method offers advantages such as faster measurements, better robustness against environmental disturbances, real-time capability, and enhanced stability \cite{Petry:20}.

The test component, as described in \ref{sssec:defect}, was subjected to thermal loading using two symmetrically arranged 400~W halogen lamps. To ensure uniform illumination of the surface, the lamps were positioned at a distance of 125~mm from the specimen support plane and spaced 140~mm apart. Measurements were taken during both heating and cooling phases, with the thermal loading applied in 1.6-second bursts starting at $t~=~0~\text{s}$, followed by 5 seconds of cooling. This cycle was repeated five times to ensure reproducibility. The thermal imaging camera FLIR A65 was used to confirm that the heat flux was uniformly distributed across the region of interest and to verify that the specimen had returned to its initial temperature state before each subsequent loading cycle. 

\section{Results and Discussion}
The comparison of experimental and simulated results focuses on both quantitative and qualitative analyses of the shearographic measurements. The main objective is to validate the simulation framework’s accuracy in predicting the optical response of the test component under thermal loading. Fringe pattern analysis is used to visually compare the fringe structures from both experimental and simulated data. This qualitative assessment examines the orientation, density, and distribution of fringes, aiming to identify both similarities and discrepancies.

\subsection{Fringe Pattern Comparison}\label{ssec:comparison}
\begin{figure}[htbp]
    \centering
    \includegraphics[width=0.8\linewidth]{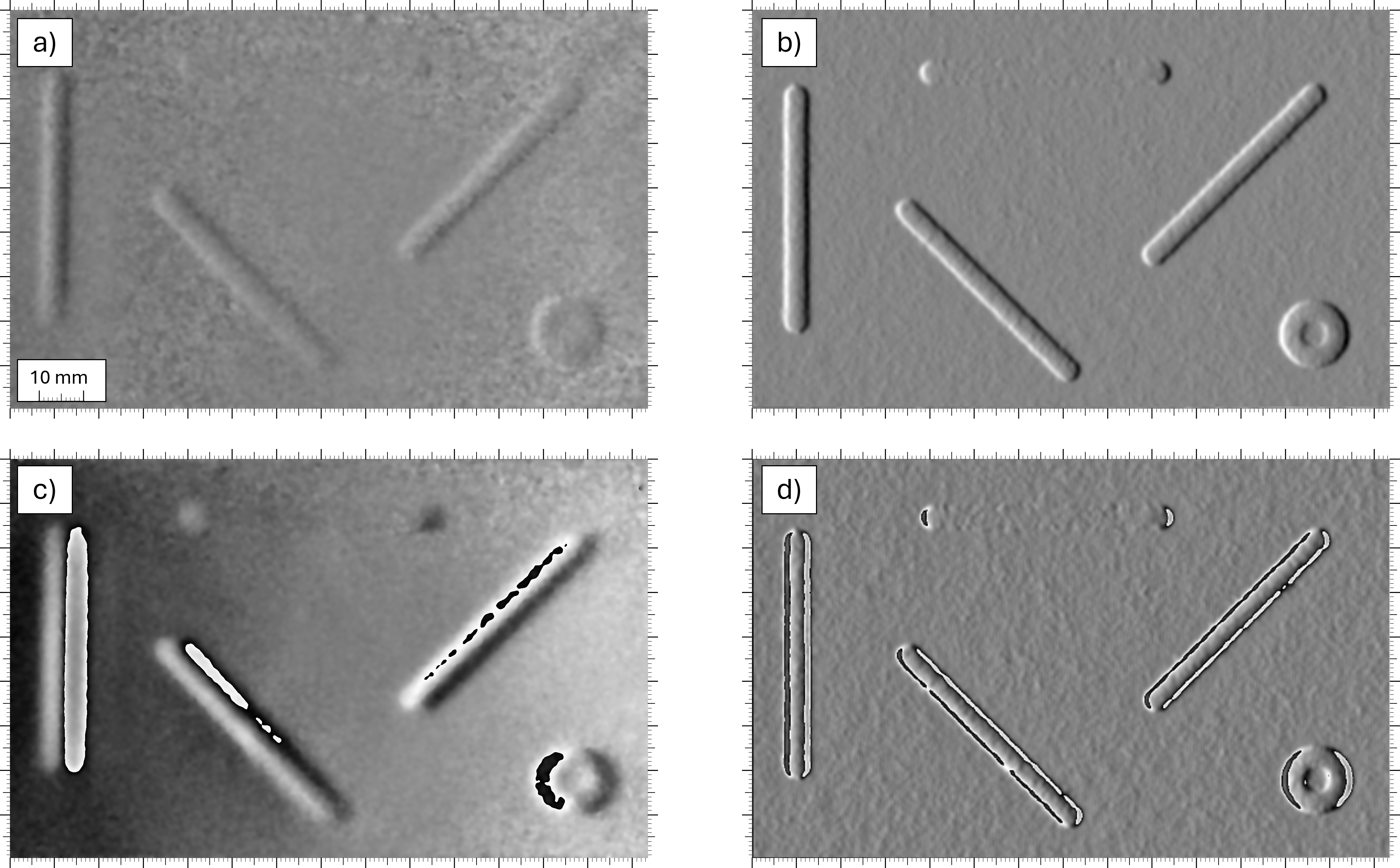}
    \hfill
    \caption{Fringe patterns from real measurements (a, c) and simulations (b, d) acquired during the cooling phase at a wavelength of 635~nm. The images show the phase difference between two deformation states recorded at \( t = 2\,\text{s} \) and \( t = 4\,\text{s} \) for shearing distances of (a, c) 1~mm and (b, d) 5~mm in the $x$-direction.}
    \label{fig:comparison_shear}
\end{figure}

To illustrate the measurement results, images acquired at a wavelength of 635~nm (red laser light) with shearing distances of 1~mm and 5~mm in the $x$-direction are compared (see Fig.~\ref{fig:comparison_shear}). The measurements were performed during the cooling phase, where global deformations exert less influence on the measurement accuracy. The first acquisition was recorded shortly after heating at \( t = 2~\text{s} \), and the second at \( t = 4~\text{s} \).

For a shearing distance of 1~mm (Fig.~\ref{fig:comparison_shear}a, b), no phase jump occurs in both the measurements and the simulation, resulting in limited interpretability despite the visibility of some of the prepared defects. In contrast, a shearing distance of 5~mm (Fig.~\ref{fig:comparison_shear}c, d) induces clear phase jumps in both the experiment and the simulation. 
For the defect aligned perpendicular to the shear direction, the phase jump is clearly visible in both the experimental and simulation results. However, for the defect aligned parallel to the shear direction, a phase jump is only observed in the simulation. In the experiment, the deformation change is detectable, but it is not sufficient to reliably identify the defect.

Furthermore, grooves oriented at $45^\circ$ to the shearing direction exhibit a weaker contrast compared to grooves perpendicular to the shearing direction, both in the experiment and the simulation. The flat-bottom hole located in the lower right corner of the specimen is also detectable. However, due to material alterations at the bore site, the center of the hole exhibits a faster relaxation behavior, resulting in an inverted butterfly-pattern within the hole's center. While this inverted deformation pattern can be anticipated at the center, it cannot be clearly detected at this wavelength in the experiment due to the absence of a distinct phase jump. In the simulation, a small phase jump clearly reveals the inverted butterfly pattern, thereby making it detectable. A more detailed analysis of the influence of material modifications on the measurement response, particularly with different illumination wavelengths, will be provided in section~\ref{ssec:wavelength}.

In the measurements, we determined the background noise, which allowed us to calculate the signal-to-noise ratio (SNR). From these measurements, we found a microstrain value of around 80~$\mu \epsilon$, with the corresponding SNR being 12~dB. In the simulation, we obtained a maximum microstrain value of 67~$\mu \epsilon$, which is lower than the experimental value. This difference is mainly due to the absence of experimental background noise. However, the simulation data still contain numerical noise from discretization, which was reduced by applying a Gaussian filter ($\sigma = 5$), resulting in an effective SNR of approximately 7~dB and enabling clearer strain evaluation. A comparison of the deformation changes reveals that the regions with a phase jump in the simulation are spatially more confined than in the experiment. This discrepancy can be attributed to filtering, measurement conditions, and the experimental setup, which broaden these regions in the measured data.

While small deformation gradients, especially at geometrical discontinuities like the edges of the upper groove, are easily detectable in the simulation, they may be obscured in the experiment by environmental factors, such as global deformations or measurement noise, which are not present under ideal simulation conditions.

\subsection{Effect of Wavelength} \label{ssec:wavelength}

In addition to the impact of the shearing distance, the illumination wavelength also affects the measurement results. This is particularly evident when analyzing the flat-bottom hole in the lower right corner of the specimen.  The material modifications at the bore site, resulting from the manufacturing process, lead to an increase in stiffness within a 5~mm diameter region at the center of the hole. Additionally, a small air bubble, with a diameter of 0.8~mm, is present within the material and reduces the remaining wall thickness in a localized area of the bore.

The wavelength influence is linked to the measurement resolution: shorter wavelengths enable the detection of smaller deformation gradients, thereby enhancing sensitivity. As a result, deformation gradients that do not induce a phase jump at longer wavelengths (e.g., 635~nm) may cause detectable phase shifts at shorter wavelengths.

\begin{figure}[htbp]
    \centering
    \includegraphics[width=0.8\linewidth]{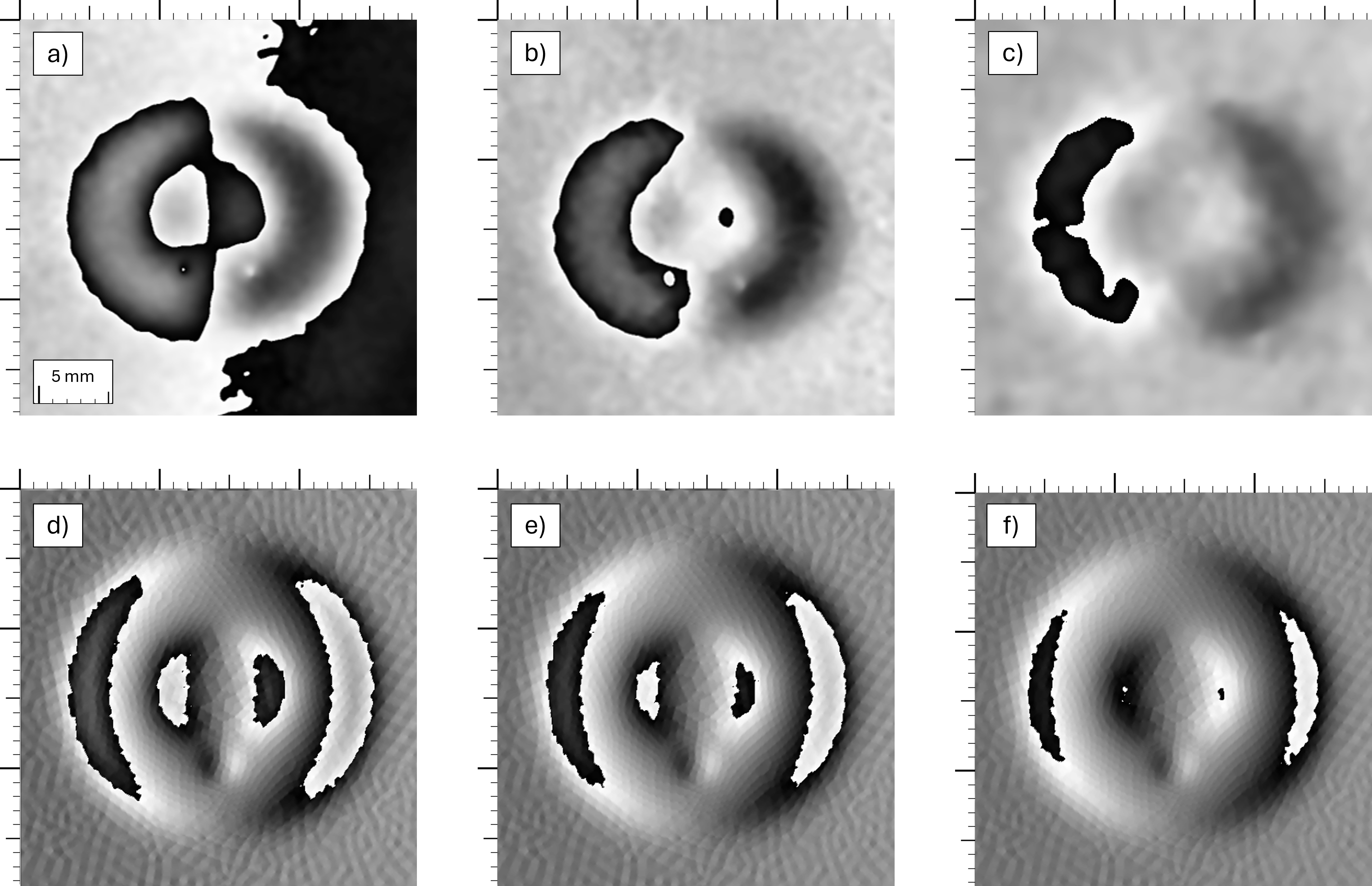}
    \caption{Measurement  and simulation results for the flat-bottom hole with a shearing distance of 5~mm at different wavelengths: (a) 450~nm, (b) 532~nm, and (c) 635~nm. The inverted butterfly pattern becomes more pronounced with shorter wavelengths.}
    \label{fig:bore_comparison}
\end{figure}

This behavior can be observed by comparing both the measurement and simulation results shown in Figure~\ref{fig:bore_comparison}, which were acquired with illumination wavelengths of 450~nm (blue), 532~nm (green), and 635~nm  (red). In the measurements, the inverted butterfly pattern is not detectable at 635~nm due to the absence of a phase jump in the center of the bore. At 532~nm, deformation indications become apparent in the measurement, and at 450~nm, the inverted butterfly pattern is clearly resolved. 

In contrast, the simulation results show a slightly stronger deformation response overall. This leads to a visible phase jump even at 635~nm, which can be attributed to the absence of experimental disturbances and to idealized material properties in the model. In reality, fluctuations in the material composition, which are common in polymer manufacturing, prevent an exact reproduction of the experimental conditions.

When comparing the shape of the outer deformation patterns, which resemble an annular or toroidal bulge, between simulation and measurement, differences become apparent. In the simulation, the deformation in $x$-direction vanishes near the center of the bore, corresponding to a region with no or very small gradient in $x$. This occurs along a vertical line at the center of the defect and is a direct consequence of the geometry, which causes the deformation to remain nearly constant in this area. Consequently, the phase jump must converge into a sharp tip. This behavior is similarly reflected in the measured data for the green and red wavelengths. 

In contrast, the measurement with blue illumination reveals a continuous phase line. This deviation can be attributed to the complex material response and the global deformation of the plate, as indicated by the vertically oriented phase ridge. Additional influences arise from the characteristics of the measurement technique, including customized filtering strategies and specific imaging settings, which affect the representation of local features. These effects are further amplified by the particular sensitivity of the measurement setup to strong shear gradients, especially in combination with the applied shear distance of 5~mm. Despite these experimental constraints, the simulated shearograms reliably reproduce the essential deformation characteristics of the complex toroidal defect. This demonstrates the capability of the simulation framework to model geometrically intricate and non-ideal cases beyond the simplified conditions considered in previous studies.

Furthermore, in the lower region of the flat-bottom hole, the simulation again exhibits more distinct deformation features compared to the measurements. According to \cite{Yang:23}, the air bubble should be difficult to detect, as the shearing amount is more than twice its maximum size. Nevertheless, the bubble is still perceptible in the measurements. However, the result is not conclusive; it merely suggests an anomaly in that region. The presence of the bubble may also contribute to the observed asymmetry in the measurements, particularly at green and blue wavelengths, where the deformation pattern deviates from mirror symmetry.

\section{Conclusion}
The simulations closely reproduce the experimental data and have proven instrumental in predicting the measurement parameters that are necessary for the successful detection of complex defects. Parameters such as shearing distance and illumination wavelength could be determined in advance, allowing for a targeted configuration of the measurement setup. This approach eliminates the need for time-consuming and potentially risky trial-and-error procedures, particularly when testing sensitive components.

Moreover, the simulation results provide insight into the sensitivity limits of the measurement system. Under idealized conditions, the simulated deformation amplitudes are more pronounced than those observed in the experiments. This enhanced sensitivity enables the detection of subtle deformation gradients that may remain hidden in real measurements due to noise, filtering effects, or material inhomogeneities. While this highlights the need for further calibration of the simulation model, it also demonstrates its value as a tool for optimizing experimental configurations and assessing the theoretical detectability of specific defect types.

In future work, the simulation data will be used to generate training datasets for the development of machine learning models aimed at defect detection. These models will incorporate the effects of different filtering strategies and aim to automate the evaluation of measurement data. The long-term goal is to implement a robust system for automated defect identification in series production, ensuring reliable and consistent quality control.

\begin{backmatter}
\bmsection{Funding}
This research was funded by the Ministry of Science and Health Rhineland-Palatinate as part of the Young Researchers Fund of the Research Initiative (Funding Period 2025) and by the Federal Ministry for Economic Affairs and Climate Protection (BMWK) through the ‘Central Innovation Program for SMEs (ZIM),’ funding code KK5060010SY3.

\bmsection{Acknowledgment}
The authors would like to thank the Ministry of Science and Health Rhineland-Palatinate and Trier University of Applied Sciences for supporting the publication of this research through the Young Researchers Fund. We also acknowledge the BMWK's ZIM program, which made the employment of researchers such as Ms. Plassmann possible. Furthermore, we extend our appreciation to Tenta Vision GmbH for providing additional equipment, such as their Sensors Agelis Terra, Flora and Aqua, which complemented our own resources and helped strengthen the validity of our simulation results. We also thank our colleagues for their valuable feedback and support during this research.

\bmsection{Disclosures}
The authors declare no conflict of interest.

\bmsection{Data availability}
Data underlying the results presented in this paper are not publicly available at this time but may be obtained from the authors upon reasonable request.

\end{backmatter}

\bibliography{literature}

\end{document}